\documentclass{aa}
\usepackage{graphicx}
\usepackage{times}
\usepackage{mathptm}
\fontfamily{ptmcm}\selectfont
\begin{document}

%\thesaurus{04.19.1; 13.18.1; 11.05.2 }
	   
\title{The ATESP radio survey}

\subtitle{III. Source counts}

\author{I. Prandoni \inst{1,2}
	\and L. Gregorini \inst{3,2}
	\and P. Parma \inst{2}
	\and H.R. de Ruiter \inst{4,2}
	\and G. Vettolani \inst{2}
	\and M.H. Wieringa \inst{5}
	\and R.D. Ekers \inst{5}
	}
\offprints{I. Prandoni}
\mail{prandoni@ira.bo.cnr.it}

\institute{Dipartimento di Astronomia, Universit\`a di Bologna, via Ranzani 1,
I--40126, Bologna, Italy
\and Istituto di Radioastronomia, CNR, Via\,Gobetti 101, I--40129, 
Bologna, Italy 
\and Dipartimento di Fisica, Universit\`a di Bologna, Via Irnerio 46,
I--40126, Bologna, Italy
\and Osservatorio Astronomico di Bologna, Via Ranzani 1, I--40126, 
Bologna, Italy
\and Australia Telescope National Facility, CSIRO, P.O. Box 76, Epping, 
NSW2121, Australia
}

\date{Received 04 August 2000 / Accepted 16 October 2000}

\titlerunning{The ATESP radio survey. III}
\authorrunning{I. Prandoni et al.}

\abstract{
This paper is part of a series reporting the results 
of the ATESP radio survey obtained at 1.4~GHz with the Australia Telescope 
Compact Array.
The survey consists of 16 radio mosaics with $\sim 8\arcsec \times 14\arcsec$
resolution and uniform sensitivity ($1 \sigma$ noise 
level $\sim$79~$\mu$Jy) over the whole area of the ESO Slice Project 
redshift survey ($\sim 26$~sq.~degrees at $\delta \sim -40\degr$). 
The ATESP survey has 
produced a catalogue of 2960 radio sources down to a flux limit ($6\sigma$)
of $\sim 0.5$ mJy. \\
\indent 
In this paper we present the 1.4 GHz $\log N - \log S$ relation derived from 
the ATESP radio source catalogue. The possible causes of incompleteness  
at the faint end of the source counts are extensively discussed and their 
effects are quantified and corrected for. The ATESP counts are 
consistent with others reported in the literature, even though some 
significant discrepancies are present at low fluxes (below a few mJy). 
We investigate whether such discrepancies may be explained in terms of 
field-to-field anisotropies, considering the fact 
that all the existing sub-mJy surveys cover small areas of sky
(from a fraction of square degree to a few square degrees). 
We stress that the ATESP survey, covering
26~sq.~degrees, provides the best determination of source counts
available today in the flux range $0.7\la S_{1.4\;\rm{GHz}}\la 2$ mJy.
}

\maketitle

\section{Introduction}\label{sec-intr}

The statistical study of radio sources allows us to address a 
number of astrophysical and cosmological problems, ranging from the 
evolution of classical radio galaxies and quasars to the properties of 
starburst galaxies as a function of cosmic time. 
The radio source counts are the most immediate product which can be
derived from a radio survey and reflect
the statistical properties of the radio source populations. 
The slope of the counts is determined essentially by the relative 
contribution of different types of sources at every flux, which is the 
result of their luminosity functions at various redshifts. 
The source counts therefore represent the most immediate observational
constraint to evolutionary models of radio sources.
An accurate determination of the source counts, their slope and normalization,
is however necessary in order to make this constraint really useful. 
The source counts, together with measured local radio luminosity functions, 
can then be used to make predictions 
(about the redshift distribution, for instance) which can be 
verified once an extensive program of optical identification and 
spectroscopy has been completed in some specific flux range. \\ 
\indent 
Normalized differential source counts derived from deep 1.4 GHz surveys 
show a flattening below a few mJy. This change in slope is 
usually interpreted as the result of the emergence of a new population  
which does not appear at higher flux densities, where the counts are
dominated by the classical powerful radio galaxies 
and quasars ($99\%$ above 60 mJy, Windhorst et al. 1990). 
This new population is thought to be a mixture
of several different types of objects (faint AGNs, normal spirals and 
ellipticals, starburst galaxies, etc.) whose relative importance 
changes as a function of the flux itself. 
Unfortunately the existing samples of faint radio sources are confined to 
regions of relatively small angular sizes and no complete optical
follow--up of such samples has been carried out yet (see {\it e.g.} Gregorini 
\& Prandoni 2000 for a review). It follows that conclusions about the 
composition of the faint radio sources are based on small and only 
partially identified samples, biased by the fact that only the brightest 
optical counterparts have spectral information. 
Many questions are therefore still open about the true nature and 
evolution of the faint radio source population and very little is 
known about the luminosity properties and redshift distribution of such 
sources. \\
\indent
The ATESP 1.4 GHz survey (Prandoni et al. 2000a, paper I) - carried 
out with the Australia Telescope Compact Array in the past years -
consists of 16 radio mosaics with $\sim 8\arcsec \times 14\arcsec$
resolution and uniform sensitivity ($1 \sigma$ noise 
level ranging from 69 $\mu$ Jy to 88 $\mu$Jy depending on the radio mosaic)
covering a narrow strip of $26\degr \times 1\degr$
at $\delta \sim -40\degr$. The ATESP survey has 
produced a catalogue of 3172 radio components, corresponding to
2960 distinct radio sources, down to a flux limit ($6\sigma$) of
$\sim 0.5$ mJy (Prandoni et al. 2000b, Paper II). 
The ATESP sample represents the largest sample (26~sq.~degr.) sensitive to
sub--mJy fluxes available so far.
It offers a unique opportunity to investigate the nature and
evolution of the mJy and sub-mJy populations and  
provides a very accurate determination of the 
radio source counts down to sub-mJy fluxes 
(the most accurate in the range $0.7 - 2$ mJy). In this respect the 
ATESP survey 
fully complements the FIRST (White et al. 1997), which
provides the most accurate determination of the source counts between 2 and 30
mJy. \\

This paper is organized as follows. The catalogue completeness is discussed in 
Sect.~\ref{sec-corr}. The ATESP number counts 
are derived in Sect.~\ref{sec-counts} and the possible 
effect of clustering on the faint end of the counts  is discussed in
Sect.~\ref{sec-clust}. Results are summarized in 
Sect.~\ref{sec-summary}.  

\section{Catalogue Completeness}\label{sec-corr}

The detection of a source depends 
on the ratio of the {\it measured} peak flux density\footnote{Peak flux 
density is always intended as brightness in units of mJy per beam solid angle.}
over the {\it local} noise level. 
In order to use the ATESP source catalogue for statistical purposes
(like the derivation of the source number counts), the possible causes
of incompleteness in the catalogue have to be recognized and 
taken into account. 

\subsection{Visibility Area}

\begin{figure}[t]
\resizebox{\hsize}{!}{\includegraphics{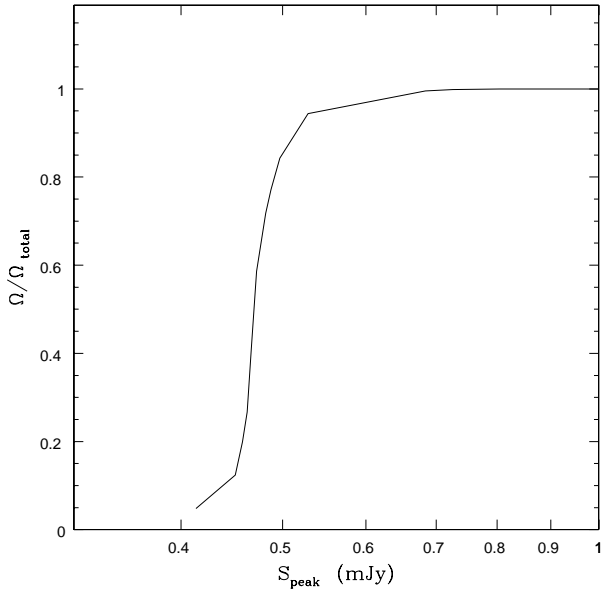}}
\caption{Visibility area of the ATESP survey. 
Fraction of the total area over which a source with given measured
peak flux density can be detected 
(i.e. $S_{\rm peak}^{\rm meas}\ge 6\sigma$). 
\label{fig-areaflux}}
\end{figure}

\begin{figure*}[t]
\resizebox{9cm}{!}{\includegraphics{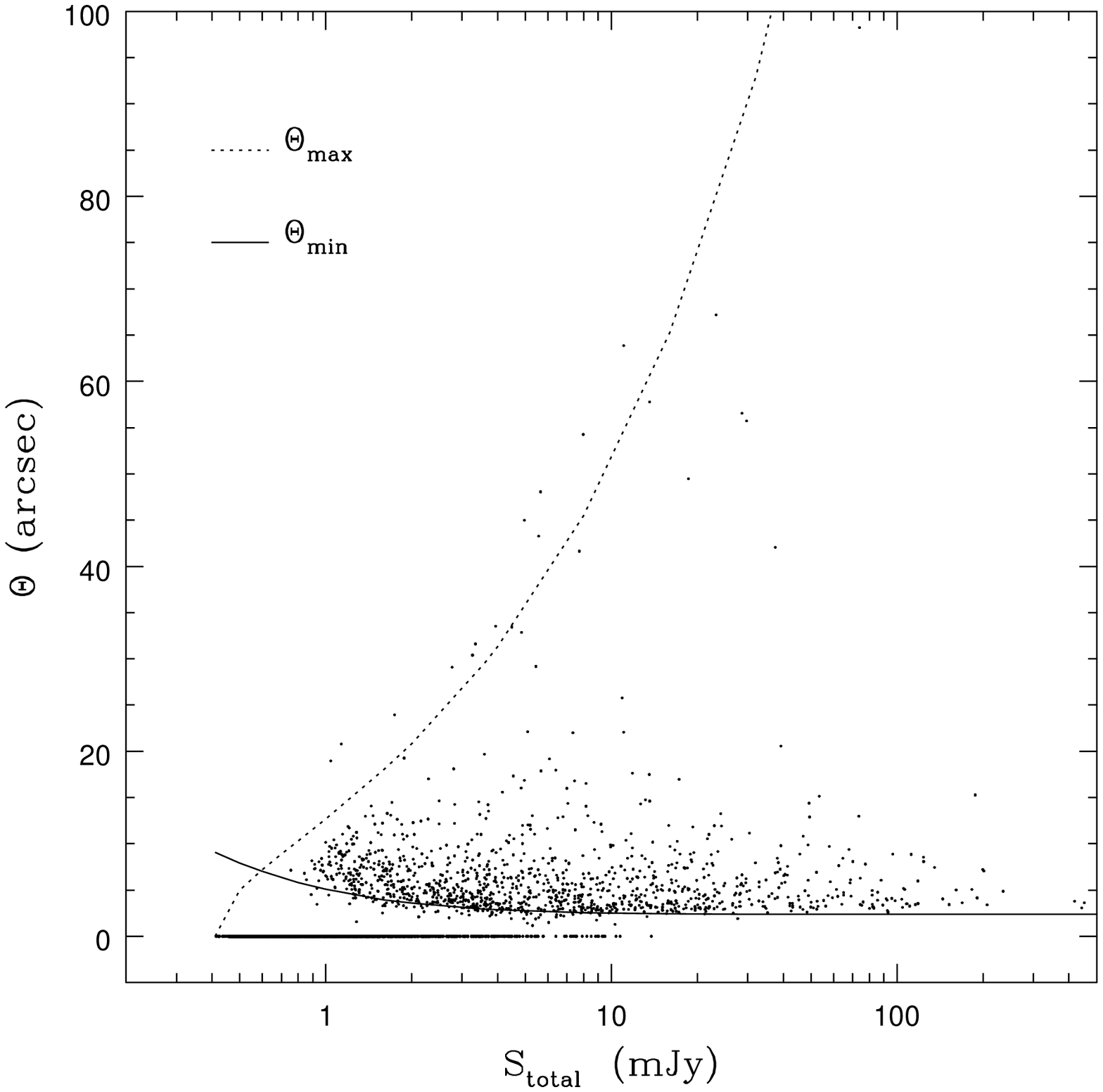}}
\resizebox{9cm}{!}{\includegraphics{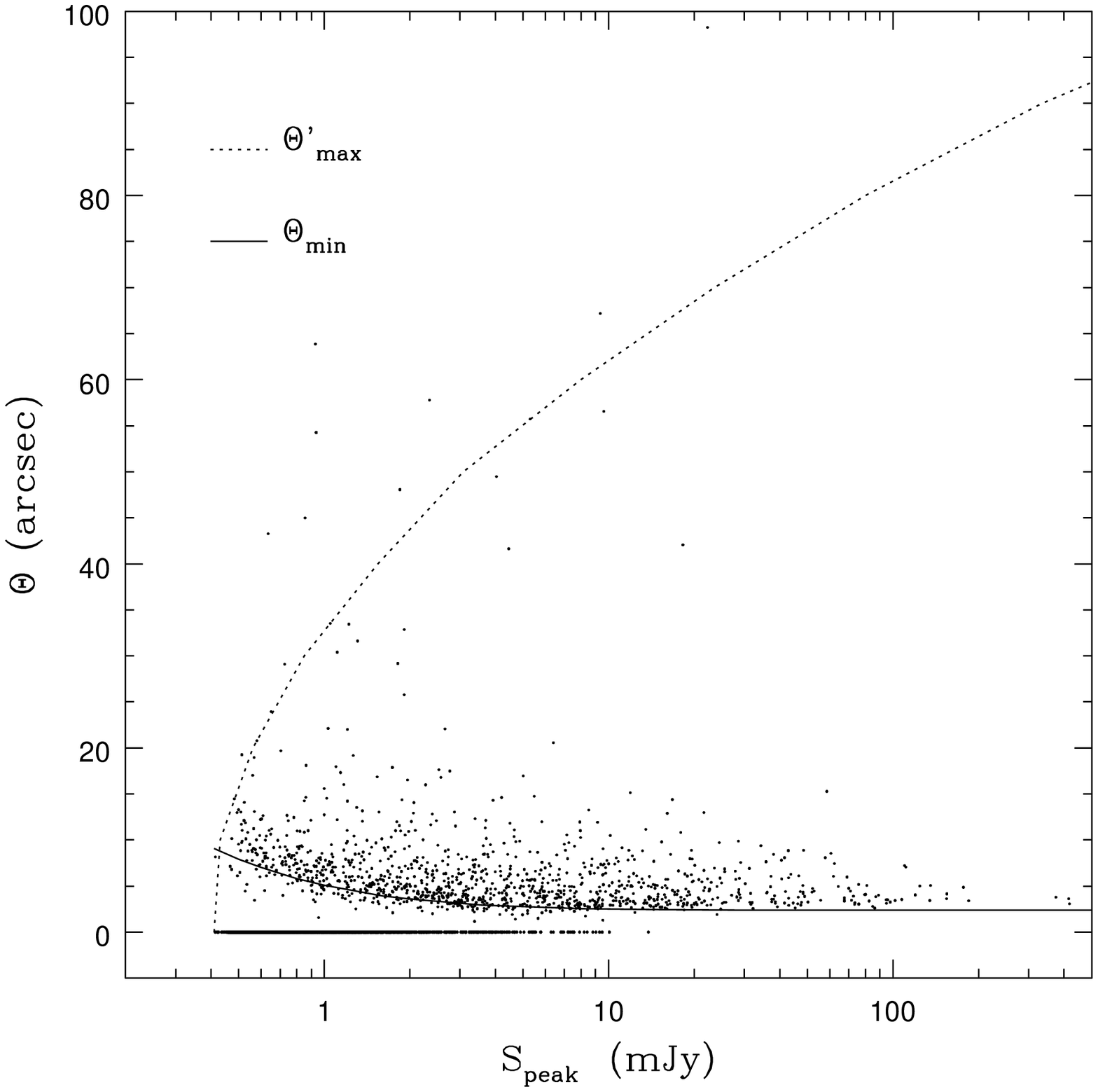}}
\caption{{\bf Left Panel.} 
Angular size (geometric mean after deconvolution) for the 3172 
ATESP radio components as a function of the measured {\em total} flux density. 
The maximum size ($\Theta_{\rm max}$) that
a source of a given $S_{\rm total}^{\rm meas}$ can have before dropping 
below the limiting flux of the survey is
indicated by the dotted line. Also indicated is 
the minimum angular size ($\Theta_{\rm min}$), below which deconvolution is
not considered significant (solid line). The lines are drawn assuming 
for the survey a $6\sigma$ 
limit of 0.41 mJy, obtained assuming the 
lowest average noise value measured in the ATESP radio mosaics (i.e. 
$\sigma=69$ $\mu$Jy, see Table~B1 in paper II). 
All unresolved sources are indicated in the figure by dots at $\Theta=0$.
{\bf Right Panel.} 
Same as in the left panel but as a function of the measured {\em peak} 
flux density. Here the dotted line indicates the maximum 
scale ($\Theta\arcmin_{\rm max}$) at which the ATESP survey is sensitive 
due to the lack of baselines shorter than
500 m, while the solid line still indicates the minimum angular 
size ($\Theta_{\rm min}$), below which deconvolution is
not considered significant. 
\label{fig-angsize}}
\end{figure*}

\begin{figure}[t]
\resizebox{\hsize}{!}{\includegraphics{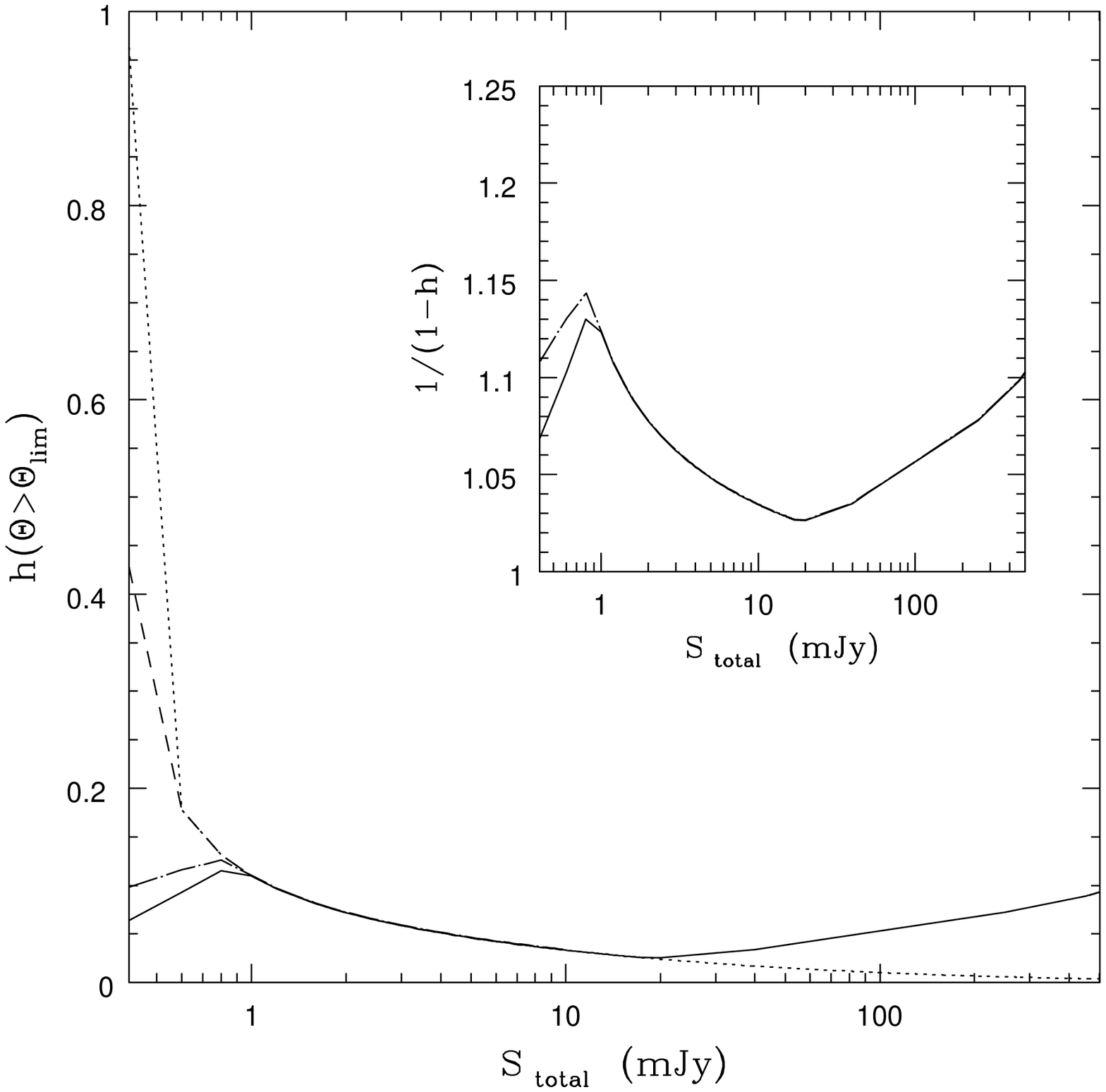}}
\caption[]{Fraction of sources with angular size larger than 
$\Theta_{\rm lim}$ ($h(>\Theta_{\rm lim})$) as a function of the total 
source flux $S_{\rm total}$ (computed as proposed by Windhorst et al. 
1990).
$\Theta_{\rm lim}$ represents the maximum size a source of given 
$S_{\rm total}$
can have in order to be detected in the ATESP survey. Different curves
are found for different definitions of $\Theta_{\rm lim}$ (see text):
$\Theta_{\rm lim}= \mbox{max} [ \Theta_{\rm min}, 
\mbox{min}(\Theta_{\rm max},\Theta\arcmin_{\rm max})]$ (solid line);
$\Theta_{\rm lim}= \mbox{max} [ 2\arcsec , 
\mbox{min}(\Theta_{\rm max},\Theta\arcmin_{\rm max})]$  (dashed line);
$\Theta_{\rm lim}=\Theta_{\rm max}$ (dotted line). Also shown is the
curve obtained by assuming $\Theta_{\rm med}=2\arcsec$ for sub-mJy sources,
as discussed in the text (dot-dashed line). The inner panel shows 
the effect of the two different definitions of $\Theta_{\rm med}$ on the
form of the resolution bias correction to be applied to 
the source counts. 
\label{fig-resbias}}
\end{figure}

Radio mosaics usually have a varying sensitivity over the 
region surveyed. This means that the area over which a source can be 
detected (visibility area, Katgert et al. 1973) increases with the source 
peak flux. 
The ATESP survey is characterized by a noise 
distribution which is fairly uniform within each mosaic and from
mosaic to mosaic ($<10\%$ variations).  
Average mosaic noise values vary from $\sim 69$ $\mu$Jy to $\sim 88$ 
$\mu$Jy (see Table~B1 in paper II) with an average 
value of $\sim 79$ $\mu$Jy ($1\sigma$). 
Also, the noise can generally be considered Gaussian, 
with only a few exceptions around strong sources ($S_{\rm peak} > 50-100$ mJy)
where higher noise levels ($\sim 100$ $\mu$Jy) can be 
occasionally found (see Paper I and II for more details). \\
The visibility area  of the ATESP survey is shown in 
Fig.~\ref{fig-areaflux}. As expected, the fraction of the total area 
over which an ATESP 
source of a given peak flux density can be detected, increases very rapidly 
between 0.4 and 0.5 mJy and becomes equal to 1 at $S_{\rm peak} \geq 0.8$ mJy. 
This is a consequence of the ATESP observing strategy 
(uniform sensitivity in the whole region surveyed) and of the fact that high 
noise regions around bright sources do not affect the source detection above 
$0.8$ mJy. 
  
\subsection{Systematic Effects}\label{sec-syst}

Two additional effects could be responsible of partial incompleteness at low 
signal-to-noise ratios: the clean bias and the bandwidth smearing. 
The clean bias has 
been extensively discussed in paper I of this series. It is responsible for
both total and peak flux density under--estimations of the order of 
10--20\% at the lowest flux levels ($S_{\rm peak}<10\sigma$) and 
gradually 
disappears going to higher fluxes (no effect for 
$S_{\rm peak}\geq 50-100\sigma$). 
The bandwidth smearing, extensively 
discussed in paper II, could allow for an extra 5\% under--estimation of the 
peak flux densities only. \\
\indent
In paper II (Appendix B) we gave a formula  (Eq.~(B1)) 
to correct the measured (peak and total) fluxes for 
the clean bias and bandwidth smearing. Such a correction varies from source
to source depending on the radio 
mosaic and on the source signal-to-noise ratio. 
However, peak flux densities are {\it at most} underestimated 
by a factor of $\sim 25\%$ due to the combined effects of clean bias and 
smearing (see paper I and II). It follows that {\it at most}
$S^{\rm meas}_{\rm peak}\sim 0.75 S^{\rm corr}_{\rm peak}$. 
This means that
any source with {\it corrected} peak flux density 
$\ga 6\sigma/0.75$ will have effective 
{\it measured} peak flux densities still above the detection threshold of the
ATESP survey.
In other words, incompleteness due to clean bias and bandwidth 
smearing does not affect the ATESP catalogue at {\it corrected} peak 
fluxes 
$> 0.55-0.70$ mJy, the precise value depending on the radio mosaic. 

\subsection{Resolution Bias}\label{sec-resb}

In deriving the source number counts, the completeness of the catalogue
in terms of total flux density is very important. 
Such a completeness depends on the source 
angular size and its component flux ratio. A resolved source of given 
$S_{\rm total}$ will drop below the $6\sigma$ peak flux density cut--off 
more easily than a point source of same $S_{\rm total}$. 
This is the so--called resolution bias. \\ 
Eq.~(1) of Paper II can be used to give an approximate 
estimate of the maximum size ($\Theta_{\rm max}$) 
a source of given $S_{\rm total}$ can have 
before dropping below the $6\sigma$ limit of the ATESP catalogue. 
In Fig.~\ref{fig-angsize} (left panel) we plot the angular size
($\Theta$) of the ATESP radio sources (or source components) as a 
function of the measured total flux density.
$\Theta$ is defined as the geometric 
mean of the (Gaussian) major and minor deconvolved axes 
of the ATESP sources as listed in the ATESP catalogue (see Paper II).
We notice that for unresolved sources (dots at $\Theta = 0$)  
we assume $S_{\rm total}=S_{\rm peak}$.
As expected, the angular sizes of the largest ATESP sources are 
in good agreement with the estimated $\Theta_{\rm max}$ (dotted line).\\
The $\Theta_{\rm max}-S_{\rm total}$ constraint is complemented by a 
second one which is a function of the peak flux and 
takes into account, through the so--called visibility function, the maximum 
scale at which the ATESP survey is sensitive due to the lack of baselines 
shorter than 500 m (see paper I, Sect.~4, for more details). 
This latter constraint ($\Theta\arcmin_{\rm max}$) is indicated 
by the dotted line in the right panel of Fig.~\ref{fig-angsize} and, as 
expected, follows the angular size distribution of the largest ATESP sources 
as a function of their measured peak fluxes. 
The $\Theta\arcmin_{\rm max}-S_{\rm peak}$ relation is important at 
high fluxes ($\ga 20$ mJy), 
where it provides a more stringent constraint than 
the $\Theta_{\rm max}-S_{\rm total}$ relation. \\
A third constraint on source angular sizes should be considered when dealing
with resolution incompleteness effects. 
As discussed in Paper II, the deconvolution efficiency depends on the
size of the ATESP survey synthesized beam ($\sim 8\arcsec \times 14\arcsec$) 
and increases with the source signal-to-noise ratio: at the lowest 
peak flux levels only sources with 
$\Theta\ga 8\arcsec -10\arcsec$ are reliably 
resolved; on the contrary, at the highest peak flux densities, sources with 
angular sizes as small as $\Theta\sim 2\arcsec$ can be resolved. 
Eqs.~(1) and (2) of 
Paper II can be used to derive an approximate estimate of the minimum 
angular size ($\Theta_{\rm min}$) that is reliably resolved as a function of 
the source peak flux. $\Theta_{\rm min}$ is indicated by the solid lines 
in Fig.~\ref{fig-angsize} 
(since $S_{\rm total}=S_{\rm peak}$ for unresolved sources, we have plotted it
in both panels). 
The $\Theta_{\rm min}$ constraint is 
important at low flux levels   
where $\Theta_{\rm max}$ and $\Theta\arcmin_{\rm max}$ become both unphysical 
(i.e. $\rightarrow 0$).\\
The three constraints discussed above ($\Theta_{\rm max}$, 
$\Theta\arcmin_{\rm max}$ and 
$\Theta_{\rm min}$) have been used to define an overall angular size 
upper limit, $\Theta_{\rm lim}$, as a function of the ATESP source flux 
density:
\begin{equation}\label{eq-thetalim}
\Theta_{\rm lim} = \mbox{max} [ \Theta_{\rm min}, 
\mbox{min}(\Theta_{\rm max},\Theta\arcmin_{\rm max})] \; .
\end{equation}
In order to estimate the incompleteness in the ATESP catalogue due to 
resolution bias, we need to make an assumption about the {\it true} angular 
size distribution of radio sources as a function of flux. 
Following Windhorst et al. (1990), we have assumed 
$h(>\Theta_{\rm lim}) = e^{-ln{2}\;(\Theta_{\rm lim}/\Theta_{\rm med})^{0.62}}$
and $\Theta_{\rm med} = 2\arcsec \cdot S_{1.4 \; \rm GHz}^{0.30}$ ($S$ in mJy).
$\Theta_{\rm med}$ represents the estimated source median angular size at a 
given flux density. 
The Windhorst et al. equations together with Eq.~(\ref{eq-thetalim}) 
allowed us to estimate the fraction of sources of given $S_{\rm total}$ 
with angular sizes larger than $\Theta_{\rm lim}$ and thus lost by the 
ATESP survey (solid line in Fig.~\ref{fig-resbias}). 
The rising slope of the curve at high flux densities is due to the 
constraint provided by $\Theta\arcmin_{\rm max}$. On the other hand, the 
decreasing trend at low fluxes is a consequence of the looser constraint 
provided by $\Theta_{\rm min}$. 
In fact, introducing $\Theta_{\rm min}$ in the equation 
takes into account the effect of having a finite synthesized beam size
(that is $\Theta_{\rm lim}>>0$ at the survey limit) and a deconvolution
efficiency which varies with the source peak flux. \\ 
In determining the resolution bias effect, a correct definition of  
$\Theta_{\rm lim}$ is very important. 
The dashed and the dotted curves in Fig.~\ref{fig-resbias} 
demonstrate how critical is the determination of the resolution bias effect 
at low signal-to-noise ratios. Slightly changing the definition of 
$\Theta_{\rm lim}$ can in fact produce very different results. 
The dashed line shows the effect on $h(>\Theta_{\rm lim})$ when a fixed value
for $\Theta_{\rm min}$ in Eq.~(\ref{eq-thetalim}) in assumed. In this case  
$\Theta_{\rm min}$ was set equal to $2\arcsec$, that is the minimum angular 
size that is reliably deconvolved at the highest flux densities, in the 
case of the ATESP survey. 
The dotted line shows, on the contrary, the effect of letting 
$\Theta_{\rm lim}\rightarrow 0$ for 
$S\rightarrow 6\sigma$, by defining $\Theta_{\rm lim}=\Theta_{\rm max}$
(we also neglect the effect of $\Theta\arcmin_{\rm max}$). The fraction
of lost sources will approach 1 at the survey limit (dotted line), 
while it monotonically decreases going to higher fluxes. \\
We have also investigated the effect of assuming a constant value  
$\Theta_{\rm med}=2\arcsec$ for sources with $S\leq 1$ mJy, as suggested by 
recent studies (e.g. Windhorst et al. 1993, Richards
2000). The corresponding $h(>\Theta_{\rm lim})$ is indicated by the dot-dashed
line in Fig.~\ref{fig-resbias}. As shown in the figure,  
this more realistic assumption results in a larger  
fraction of lost sources at $S<1$ mJy ($10\%$ instead of $6\%$ at the survey 
limit). The uncertainty in the form of $\Theta_{\rm med}$ at sub-mJy fluxes
obviously translates in an uncertainty in the form of the resolution bias 
correction ($c=1/[1-h(>\Theta_{\rm lim})]$) to be applied to the source 
counts (see inner panel of Fig.~\ref{fig-resbias}).
At sub-mJy fluxes, 
this upward systematic uncertainty ($\leq 4\%$) has to be (quadratically) 
added to a more general 10\% uncertainty in the resolution bias correction 
(as indicatively quantified by Windhorst et al. 1990). 

\begin{figure*}[t]
\resizebox{10.5cm}{!}{\includegraphics{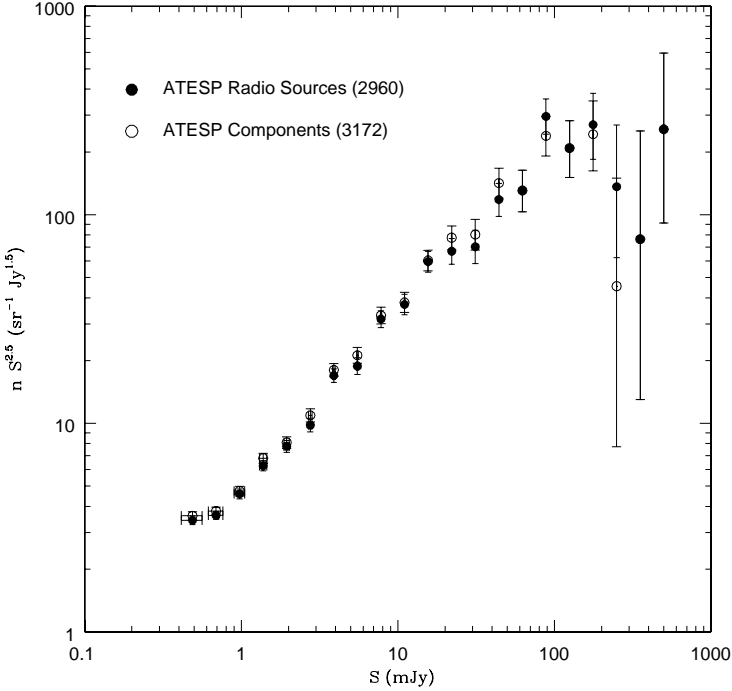}}
\hfill
\parbox[b]{55mm}{
\caption{Normalized 1.4 GHz differential source counts derived
from the ATESP catalogue. Filled circles refer to the ATESP radio source 
catalogue (2960 distinct objects). Empty circles refer to the list of all the 
ATESP components (3172 entries). Horizontal bars represent $1\sigma$
errors in the {\it measured} total flux. Vertical bars
represent Poissonian errors.}
\label{fig-counts.atesp}}
\end{figure*}

\begin{figure*}[]
\resizebox{10.5cm}{!}{\includegraphics{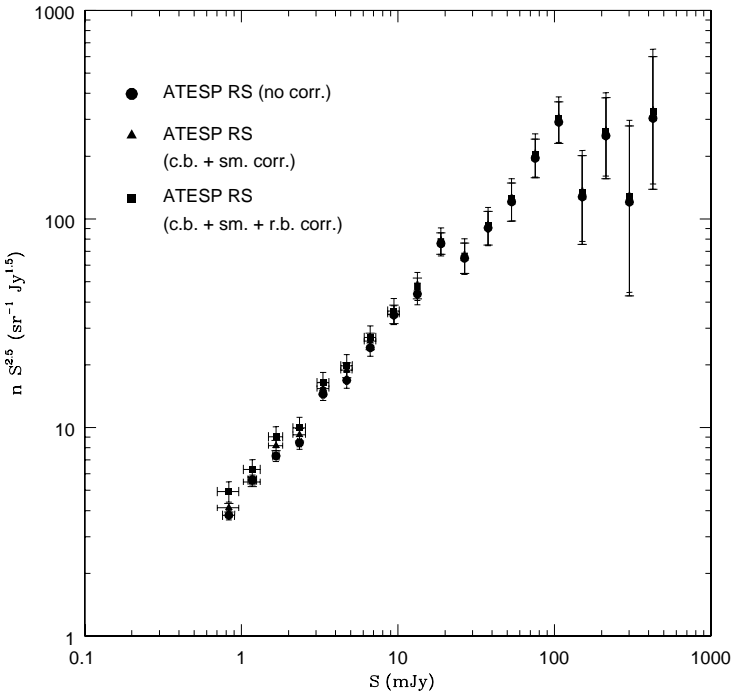}}
\hfill
\parbox[b]{55mm}{
\caption{Normalized 1.4 GHz differential source counts derived  
from the ATESP catalogue (multiple sources are counted as one object). 
The source counts are plotted before 
(filled circles, same as shown in Fig.~\ref{fig-counts.atesp}) and after 
correcting for clean bias and bandwidth smearing (filled triangles). 
In the latter case horizontal bars represent $1\sigma$
errors in the {\it corrected} total flux.
Filled squares represent the counts 
obtained once the resolution bias is also taken into account. 
Vertical bars include here the uncertainty in the resolution bias correction.
In all cases the counts
have been derived down to 0.70 mJy (instead of 0.41 mJy), in order to 
avoid the incompleteness effects introduced by the flux corrections.  }
\label{fig-counts.atesp.corr}}
\end{figure*}

\begin{figure*}[t]
\resizebox{12cm}{!}{\includegraphics{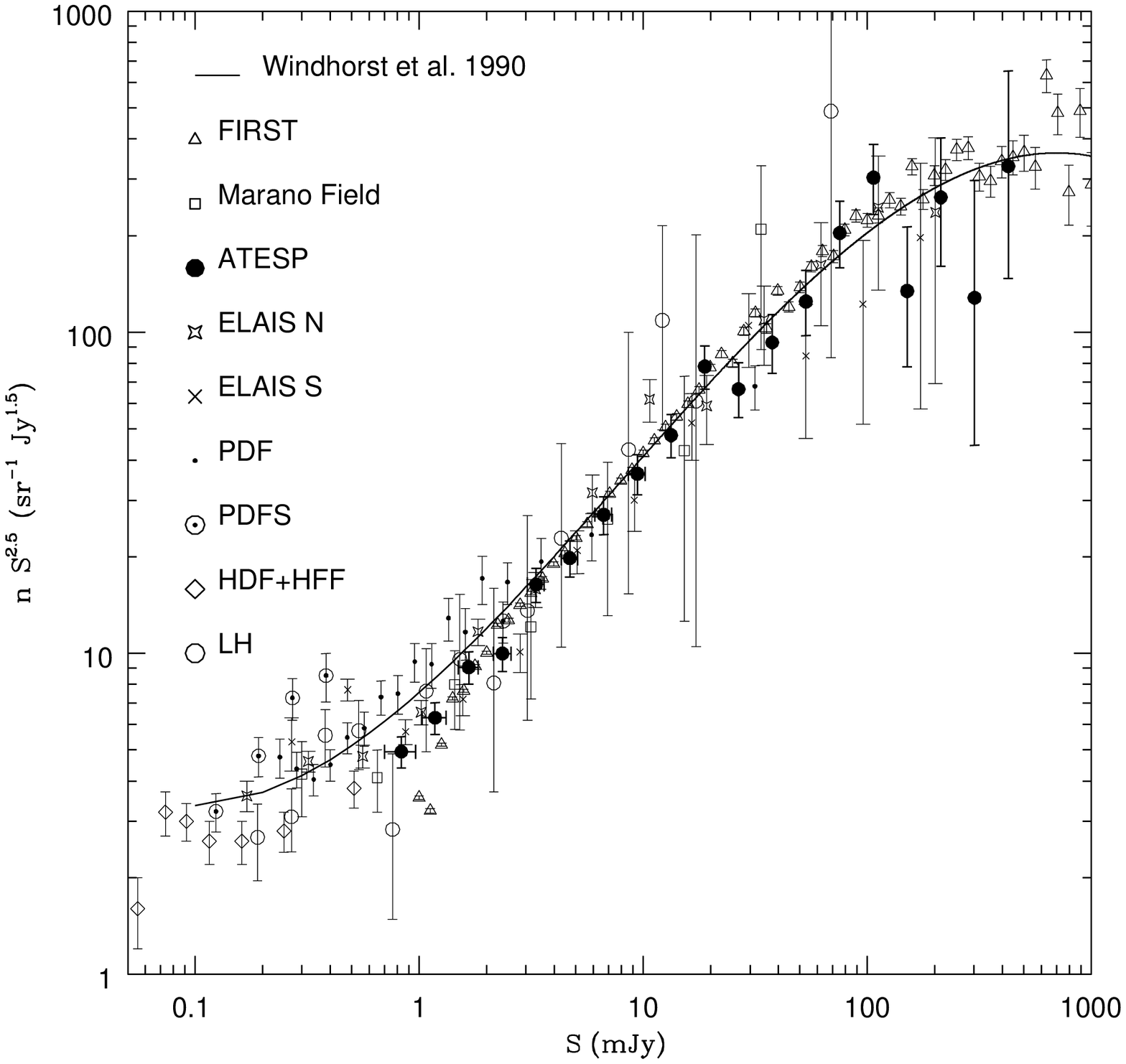}}
\hfill
\parbox[b]{55mm}{
\caption{Normalized 1.4 GHz differential source counts for 
different samples: the FIRST (triangles, White et al. 1997), the Marano Field 
(squares, Gruppioni et al. 1997), the ELAIS North and South
(stars and crosses, Ciliegi et al. 1999 and Gruppioni et al. 1999), 
the Phoenix Deep Survey (dots and circled dots, Hopkins et al. 1998), the 
Hubble Deep Field + Flanking Fields (diamonds, Richards 2000) and 
the Lockman Hole (empty circles, de Ruiter et al. 1997). 
The fit obtained by Windhorst et al. (1990) is also shown (solid line). 
The ATESP source counts presented here (filled circles), are 
corrected for clean bias, bandwidth smearing and resolution bias 
(same as filled squares presented in Fig.~\ref{fig-counts.atesp.corr}).
Horizontal bars associated to the ATESP counts represent $1\sigma$
errors in the {\it corrected} total flux, while vertical bars include 
the uncertainty in the resolution bias correction.
}
\label{fig-counts}}
\end{figure*}

\section{The ATESP Source Counts}\label{sec-counts}

We used the $6\sigma$ ATESP catalogue (Paper II) 
to derive the differential ATESP source 
counts down to 0.41 mJy. 
In computing the counts we have used the integrated flux 
for extended sources and the peak flux for point-like sources. 
Each source has been weighted for the reciprocal of its visibility area 
($\Omega(S_{\rm peak})/\Omega_{\rm total}$, see Fig.~\ref{fig-areaflux}), that is 
the area over which the source could be detected. We
notice that for $S_{\rm peak}>0.8$ mJy a source can be counted over the whole 
survey area. \\
\indent
In Fig.~\ref{fig-counts.atesp} we present the ATESP counts derived from
the 3172 ATESP components (empty circles) and from the 2960 distinct 
radio sources defined from them, that is counting 
any multi--component source as one (filled circles).
In this figure no corrections were applied to take into account 
either the systematic under--estimations of the flux densities 
(see Sect.~\ref{sec-syst})
or the resolution bias (see Sect.~\ref{sec-resb}). 
Horizontal bars represent $1\sigma$ errors in the measured flux
($\sigma(S^{\rm meas})$), calculated as discussed in Paper II.
Vertical bars represent Poissonian errors (calculated 
following Regener 1951). \\
In the ATESP catalogue a source is defined as {\it multiple} 
whenever two or more
components are found closer than $45\arcsec$ (even though some
exceptions are present, see Paper II for more details). Applying this
distance constraint we expect $\sim 20\%$ contamination by random 
superpositions. 
Fig.~\ref{fig-counts.atesp} shows how possible spurious 
multiple sources could affect the ATESP source counts. \\
In Fig.~\ref{fig-counts.atesp.corr} we show the effect of the flux and
resolution bias corrections on the ATESP source counts 
(multiple sources are again counted as one). 
Filled circles and triangles indicate the counts respectively
before and after correcting the flux densities for both the clean bias and 
the bandwidth smearing (fluxes are corrected as proposed in Paper II).
In the latter case horizontal bars represent 
$\sigma(S^{\rm corr})$, obtained by propagating the errors of the measured 
quantities $S^{meas}$, $k$, $a$ and $b$ which appear in Eq.~(B1) of Paper II.
We note that the plotted values of $\sigma(S^{corr})$ have to be considered 
as average estimates since they have been derived in the case of
{\it intermediate} clean bias correction parameters ($a=0.13\pm 0.03$ 
and $b=0.75\pm 0.06$, see discussion in Sect.~5.3 of Paper I). We also
notice that $\sigma(k)=0.006$.  
Filled squares in Fig.~\ref{fig-counts.atesp.corr} 
show the effect of adding the 
resolution bias correction (computed following Windhorst et al. 1990, as
discussed in Sect.~\ref{sec-resb}). 
In the estimation of the vertical bars we have here included 
the errors due to the uncertainty in the
resolution bias correction (see Sect.~\ref{sec-resb}). 
We notice that the ATESP counts presented in this figure 
are derived down to a limit of 0.70 mJy (instead of 0.41 mJy) in order 
to prevent the corrected counts to suffer 
from incompleteness (see discussion in Sect.~\ref{sec-corr}). \\

\begin{table}[]
\caption{ATESP source counts. 
\label{tab-diffcounts}}
\scriptsize

\begin{flushleft}
\begin{tabular}{ccrc}
\hline\hline\noalign{\smallskip}
\multicolumn{1}{c}{$\Delta S$}
& \multicolumn{1}{c}{$<S>$}
&\multicolumn{1}{c}{$N_S$}
& \multicolumn{1}{c}{$n S^{2.5}$}\\
\multicolumn{1}{c}{(mJy)}
& \multicolumn{1}{c}{(mJy)}
&\multicolumn{1}{c}{}
& \multicolumn{1}{c}{(sr$^{-1}$ Jy$^{1.5}$)}\\
\noalign{\smallskip}
\hline\noalign{\smallskip} 
0.70 -- 0.99 &  $0.83\pm 0.13 $  & 466  &    4.94$_{-0.54}^{+0.56}$ \\
\noalign{\smallskip}
0.99 -- 1.40 &   $1.18\pm 0.15$  & 372  &    6.30$_{-0.71}^{+0.72}$ \\
\noalign{\smallskip}
1.40 -- 1.98 &   $1.66\pm 0.17$  & 331  &    9.05$_{-1.03}^{+1.05}$ \\
\noalign{\smallskip}
1.98 -- 2.80 &   $2.35\pm 0.22$  & 222  &    9.99$_{-1.20}^{+1.23}$ \\
\noalign{\smallskip}
2.80 -- 3.96 &   $3.33\pm 0.29$  & 220  &    16.4$_{-2.0}^{+2.0}$ \\
\noalign{\smallskip}
3.96 -- 5.60 &   $4.71\pm 0.40$  & 160  &    19.8$_{-2.5}^{+2.6}$ \\
\noalign{\smallskip}
5.60 -- 7.92 &  $6.66\pm 0.56$ & 131  &    27.0$_{-3.6}^{+3.7}$ \\
\noalign{\smallskip}
7.92 -- 11.2 &  $9.42\pm 0.80$ & 105  &    36.2$_{-5.1}^{+5.3}$ \\
\noalign{\smallskip}
11.2 -- 15.8 & $13.3\pm  0.1$ &  83  &    47.8$_{-7.1}^{+7.5}$ \\
\noalign{\smallskip}
15.8 -- 22.4 & $18.8\pm  0.1$ &  81  &    78.2$_{-11.7}^{+12.4}$ \\
\noalign{\smallskip}
22.4 -- 31.7 & $26.6\pm 0.2$  &  41  &    66.5$_{-12.3}^{+13.7}$ \\
\noalign{\smallskip}
31.7 -- 44.8 & $37.7\pm 0.2$  &  34  &    93.0$_{-18.5}^{+20.9}$ \\
\noalign{\smallskip}
44.8 -- 63.4 & $53.3\pm 0.3$  &  27  &   125$_{-27}^{+31}$ \\
\noalign{\smallskip}
63.4 -- 89.6 & $75.3\pm 0.4$  &  26  &   204$_{-45}^{+52}$ \\
\noalign{\smallskip}
89.6 -- 127 & $107\pm  1$   &  23  &   303$_{-70}^{+82}$ \\ 
\noalign{\smallskip}
127 -- 179 &  $151\pm 1$    &   6  &   134$_{-57}^{+78}$ \\
\noalign{\smallskip}
179 -- 253 &  $213\pm  1$   &   7  &   263$_{-103}^{+140}$ \\
\noalign{\smallskip}
253 -- 358 &  $301\pm 2$    &   2  &   128$_{-84}^{+170}$ \\
\noalign{\smallskip}
358 -- 507 &  $426\pm 2 $   &   3  &   329$_{-182}^{+322}$ \\
\noalign{\smallskip}
\hline\hline
\end{tabular}
\end{flushleft}
\end{table}
\normalsize

\indent
The final ATESP source counts (those corrected for all the effects and
indicated by filled squares in Fig.~\ref{fig-counts.atesp.corr}) 
are also listed in Table~\ref{tab-diffcounts}, where,
for each flux interval ($\Delta S$), the flux geometric mean ($<S>$), 
the number of sources detected ($N_S$) and the weighted 
normalized differential counts ($nS^{2.5}$) are given. Also listed 
are the errors associated to $\Delta S$ and to the normalized counts 
(computed as discussed above). In this table fluxes are intended as
total corrected flux densities. 
The final ATESP source counts have then been compared
with the most updated previous determinations at 1.4 GHz (see 
Fig.~\ref{fig-counts}). Also shown is the 
interpolation determined by Windhorst et al. (1990) from a collection of 
$10,575$ radio sources belonging to 24 different surveys at 1.4 GHz, 
representing the state of the art at that time (solid line). 
As illustrated in Fig.~\ref{fig-counts} there is consistency between 
the ATESP counts and those obtained by 
other recent surveys, with the exception of the Phoenix Deep 
Survey (PDF and PDFS, Hopkins et al. 1998), whose counts at $S\geq 0.7$ mJy 
are systematically higher than the ATESP counts (and also higher than 
the counts derived from the other surveys presented in the figure). \\
We notice that according to the ATESP determination the upturn in the source 
counts should show up at lower fluxes ($S\la 1$ mJy)
than indicated by the Windhorst et al. (1990) fit ($S\la 5$ mJy). 
The ATESP counts are in very good agreement with the FIRST counts 
(White et al. 1997, triangles in Fig.~\ref{fig-counts}), which
are the most accurate available today over the flux range 2--30 mJy. 
Our survey, on the other hand, provides the 
best determination of the counts at fainter fluxes ($0.7<S<2$ mJy),
where the FIRST becomes incomplete. The ATESP counts can thus provide an 
useful observational constraint
on the evolutionary models for the mJy and sub--mJy populations. \\
\indent
Below a few mJy the number counts presented in Fig.~\ref{fig-counts} 
show a large spread. 
Such a spread can be due to technical reasons, like different corrections 
for resolution bias applied by different surveys ({\it e.g.} Hubble Deep 
Field, Phoenix Deep Survey, ATESP) or no correction at all (e.g. ELAIS, Marano 
Field). Note, however, that discrepancies are found even
within the same survey: the deeper PDFS counts, for instance, 
are not consistent with the PDF counts in the overlapping flux range. 
This example suggests that the effect of 
field-to-field anisotropies and/or clustering could be very important, 
since the faintest samples 
typically cover very small regions of sky ($<< 1$ sq.~degree). This latter
effect does not apply to the ATESP sample which covers 26 sq.~degrees.

\begin{figure}[t]
\resizebox{\hsize}{!}{\includegraphics{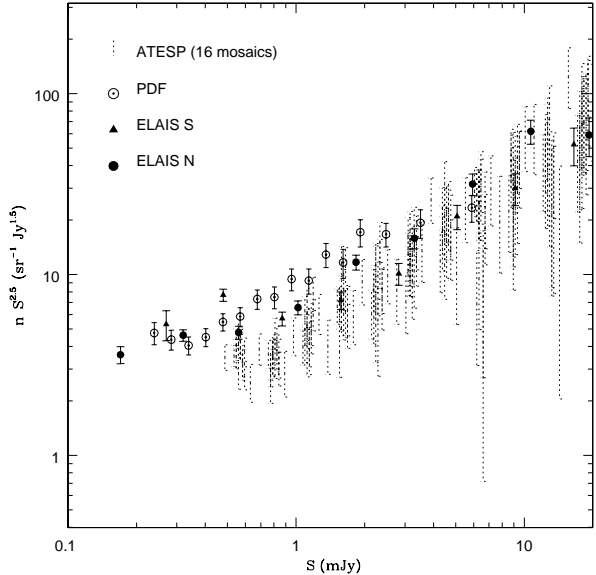}}
\caption[]{Normalized 1.4 GHz differential source counts for the 16
ATESP mosaics separately (dotted $1\sigma$ Poissonian error bars). 
For consistency, these counts are 
compared only to the ones obtained from the surveys covering areas of the 
same order of magnitude (a few sq. degrees): the ELAIS North and South
(filled circles and triangles) and the PDF (circled dots). 
The ATESP counts are not corrected for clean bias, bandwidth 
smearing and resolution bias (Horizontal error 
bars have not been drawn to 
prevent overcrowding).
\label{fig-counts.mosaic}}
\end{figure}

\begin{figure}[t]
\resizebox{\hsize}{!}{\includegraphics{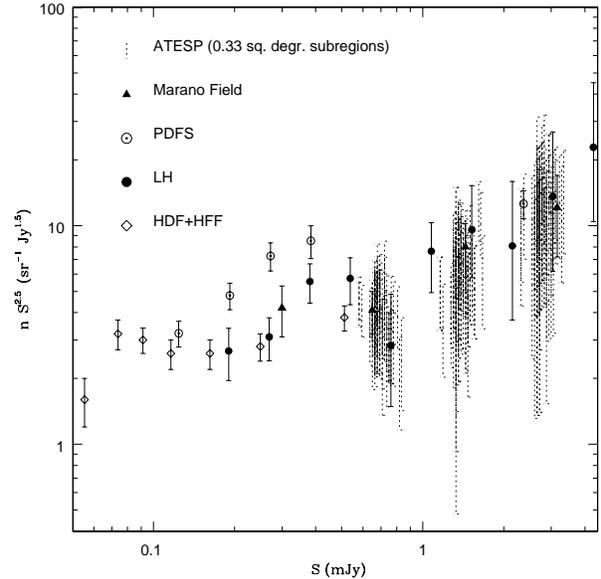}}
\caption{Normalized 1.4 GHz differential source counts for 0.33
sq. degree ATESP sub--regions (dotted $1\sigma$ Poissonian error bars). 
For consistency these 
counts are compared only to the ones obtained from the surveys covering 
areas of the same order of magnitude (a fraction of sq. degree): 
the PDFS (circled dots), the Marano Field (triangles), 
the HDF+HFF (diamonds) and the Lockman Hole (filled circles).  
The ATESP counts are not corrected for clean bias, bandwidth 
smearing and resolution bias (Horizontal error bars have not been drawn to 
prevent overcrowding).
\label{fig-counts.033}}
\end{figure}

\section{Clustering Effects on Deep Radio Counts}\label{sec-clust}

The ATESP survey allows us to test whether field-to-field anisotropies 
and/or clustering effects could explain, at least in part, 
the existing spread in source counts  at low fluxes ($S\la 1$ mJy). 
The largest sub--mJy surveys cover 
areas of the order of a few square degrees (typically 3 sq.~degr.), while 
the deepest sub--mJy surveys cover only a fraction of square degree 
(typically 0.33 sq.~degr.). We have therefore tested these two 
typical scales for sub--mJy radio surveys. The larger scale has been tested 
by deriving 
the counts for each of the 16 ATESP mosaics separately (mosaic areas range 
from 1.3 to 2 sq. degrees, see paper I). The smaller scale, on the other hand,
has been tested by splitting the ATESP survey 
in $40\arcmin \times 30\arcmin$ (0.33 sq. degrees) sub--regions. 
Note that no correction for clean bias, 
bandwidth smearing and resolution bias has been applied in this analysis 
in order to derive the ATESP counts down to the detection limit
of the survey ($6\sigma$) and to have a more meaningful comparison with 
the fainter surveys. \\
\indent
The results are reported in Figs.~\ref{fig-counts.mosaic} 
and~\ref{fig-counts.033} (dotted error bars).
For consistency, the ATESP counts obtained on these two scales 
have been compared only to surveys covering areas of the same order 
of magnitudes: 
the ELAIS and PDF for the larger scale (see Fig.~\ref{fig-counts.mosaic}) ; 
the MF, LH, PDFS and HDF+HFF for the smaller scale 
(see Fig.~\ref{fig-counts.033}). 
In either case the ATESP counts are in good agreement with the ones derived 
from the comparison samples, at least down to the fluxes probed by the ATESP. 
This is particularly true if we consider that no corrections for clean 
bias, bandwidth smearing or resolution bias have been applied to the ATESP 
counts. It is therefore probable that field-to-field anisotropies do not 
play an important role in these samples. 
The only exception is represented by the PDF counts 
(circled dots in Fig.~\ref{fig-counts.mosaic}), which remain 
systematically in excess 
(in particular at fluxes $0.5\la S\la 1$ mJy). 
On the other hand, the excess shown by the PDFS counts 
(circled dots in Fig.~\ref{fig-counts.033}) cannot be directly probed by the 
ATESP survey. 
From the analysis of Fig.~\ref{fig-counts.mosaic} it seems 
very hard to explain the PDF counts in terms of Poissonian fluctuations. 
We have applied the Kolmogorov-Smirnov 2-sample test to the PDF and the ATESP 
counts shown if Fig.~\ref{fig-counts.mosaic} (we have considered the counts 
derived from the 16 ATESP mosaics as a unique larger sample) and 
we found that the probability of the two samples being drawn from the same 
parent population is 0.2\%. This leads us to conclude that 
either the PDF sample has a significant excess of sources or some other 
unknown systematic effect is present.  
It is however worth to notice that in general systematic discrepancies like 
the ones shown by the counts derived from the Phoenix Deep Survey cannot 
be explained in terms of different assumptions for resolution bias, 
because such an effect does strongly affect the counts only at  
low signal-to-noise ratios (see discussion in Sect.~\ref{sec-resb}). 

\section{Summary}\label{sec-summary}

We have presented the $\log{N} - \log{S}$ relation derived from the ATESP 
catalogue. Particular emphasis has been given to the possible causes of 
incompleteness at the faint end of the source counts. In particular we have
discussed the effect of systematic under-estimations in the measured 
flux densities due to clean bias and bandwidth smearing and we have 
quantified the incompleteness of the catalogue in terms of total flux
density (the so--called resolution bias). 
Once corrected for such effects, the ATESP counts are consistent with
the counts derived from other recent surveys. The only discrepant 
counts are those obtained from the Phoenix Deep Survey 
(Hopkins et al. 1998), which are systematically higher. 
It is worth to notice that the ATESP counts point towards lower 
fluxes ($S\la 1$ mJy) for the upturn than those ($S\la 5$ mJy) indicated by 
the Windhorst et al. (1990) best fit.  \\
\indent
At fluxes below 1 mJy, the 1.4 GHz number counts derived from different 
surveys show a large spread. 
Such a spread can be due to technical reasons, like different corrections 
for resolution bias or no correction at all.   
The effect of field-to-field anisotropies and/or clustering could also 
play an important 
role, since the faintest samples typically cover very small regions of sky 
($<< 1$ sq.~degree). This latter
effect does not apply to the ATESP sample which covers 26 sq.~degrees. 
We have investigated such a possibility by splitting the ATESP catalogue in 
smaller sub--samples. In particular we have tested two typical scales 
for sub--mJy radio surveys: a few~sq.~degrees for the largest
surveys ($1-2$~sq.~degr. in this case) and $\sim 0.33$~sq.~degrees for the 
deepest surveys. 
As a result we have found that on both scales the spread in the counts can be 
simply explained in terms of Poissonian fluctuations, at least down to the
flux densities probed by the ATESP. 
The only exception is represented by the Phoenix Deep Survey 
whose counts remain systematically higher than the ATESP counts.  
Such a discrepancy is statistically relevant and draws us to conclude that
clustering and/or some other unknown systematic effect should play an important
role in the case of the PDF sample. \\
\indent
The ATESP survey provides the best determination of the 1.4 GHz 
source counts in the flux range 0.7--2 mJy, and in this respect it
complements the FIRST (White et al. 1997) which provides the most accurate 
counts available at higher fluxes (2--30 mJy). 
The ATESP source counts can thus be used to set 
observational constraints on the evolutionary models for the mJy and 
sub--mJy populations. 

\begin{acknowledgements}
The Australia Telescope is funded by the Commonwealth of Australia for 
operation as a National Facility managed by CSIRO.
The authors acknowledge R. Fanti for reading and commenting on an 
earlier version of this manuscript. 

\end{acknowledgements}

\end{document}